\documentclass[10pt,letterpaper]{article}
\usepackage[top=0.85in,left=2.75in,footskip=0.75in]{geometry}

\usepackage{amsmath, amssymb, amsthm, graphicx, caption, mathtools, physics, mathbbol, cancel, bm, mathrsfs}

\usepackage{changepage}

\usepackage[utf8x]{inputenc}

\usepackage{textcomp,marvosym}

\usepackage{cite}

\usepackage{nameref,hyperref}

\usepackage[right]{lineno}

\usepackage{microtype}
\DisableLigatures[f]{encoding = *, family = * }

\usepackage[table]{xcolor}

\usepackage{array}

\usepackage{url}            % simple URL typesetting
\usepackage{amsfonts}       % blackboard math symbols
\usepackage{nicefrac}       % compact symbols for 1/2, etc.
\usepackage{graphicx}
\usepackage{tikz}
\usepackage{relsize}
\usepackage{csquotes}
\usepackage{bbold}

\newcolumntype{+}{!{\vrule width 2pt}}

\newlength\savedwidth

\raggedright
\usepackage[aboveskip=1pt,labelfont=bf,labelsep=period,justification=raggedright,singlelinecheck=off]{caption}

\makeatletter
\renewcommand{\@biblabel}[1]{\quad#1.}
\makeatother

\usepackage{lastpage,fancyhdr,graphicx}
\usepackage{epstopdf}
\fancyheadoffset[L]{2.25in}
\fancyfootoffset[L]{2.25in}
\def\prl{Phys. Rev. Lett.}
\def\aap{Astron. \& Astrophys.}

\usetikzlibrary{positioning,shapes}
\usetikzlibrary{arrows,automata,shadows}
\usetikzlibrary{positioning}
\usetikzlibrary{calc}
\usetikzlibrary{decorations.pathreplacing} % For the braces
\usetikzlibrary{external}

\tikzset{
   line label/.style = {
       fill=white,
       inner sep=2pt,
       font=\small\sffamily
   },
   fontscale/.style = {font=\relsize{#1}}
}
\pgfdeclarelayer{bg}    % declare background layer
\pgfsetlayers{bg,main}  % set the order of the layers (main is the standard layer)

\begin{document}
\vspace*{0.2in}

% Title must be 250 characters or less.
\begin{flushleft}
{\Large
\textbf\newline{Causal, Bayesian, \& Non-parametric Modeling of the SARS-CoV-2 Viral Load Distribution vs.~Patient's Age}
% Please use "sentence case" for title and headings (capitalize only the first word in a title (or heading), the first word in a subtitle (or subheading), and any proper nouns).
}
\newline
% Insert author names, affiliations and corresponding author email (do not include titles, positions, or degrees).
\\
Matteo Guardiani\textsuperscript{1,2,*},
Philipp Frank\textsuperscript{1,2},
Andrija Kostić\textsuperscript{1,2},
Gordian Edenhofer\textsuperscript{1,2},
Jakob Roth\textsuperscript{1,2},
Berit Uhlmann\textsuperscript{4},
and Torsten Enßlin\textsuperscript{1,2,3}
\\
\bigskip
\textbf{1} Max Planck Institute for Astrophysics, Garching, Germany
\\
\textbf{2} Fakultät für Physik, Ludwig-Maximilians-Universität München, Munich, Germany
\\
\textbf{3} Excellence Cluster ORIGINS, Garching, Germany
\\
\textbf{4} Süddeutsche Zeitung, Munich, Germany
\\
\bigskip

% Insert additional author notes using the symbols described below. Insert symbol callouts after author names as necessary.
% 
% Remove or comment out the author notes below if they aren't used.
%
% Primary Equal Contribution Note
% \Yinyang These authors contributed equally to this work.

% Additional Equal Contribution Note
% Also use this double-dagger symbol for special authorship notes, such as senior authorship.
% \ddag These authors also contributed equally to this work.

% Current address notes
% \textcurrency Current Address: Dept/Program/Center, Institution Name, City, State, Country % change symbol to "\textcurrency a" if more than one current address note
% \textcurrency b Insert second current address 
% \textcurrency c Insert third current address

% Deceased author note
% \dag Deceased

% Group/Consortium Author Note
% \textpilcrow Membership list can be found in the Acknowledgments section.

% Use the asterisk to denote corresponding authorship and provide email address in note below.
* matteani@mpa-garching.mpg.de

\end{flushleft}
% Please keep the abstract below 300 words
\section*{Abstract}
    The viral load of patients infected with SARS-CoV-2 varies on logarithmic
    scales and possibly with age. Controversial claims have been made
    in the literature regarding whether the viral load distribution actually
    depends on the age of the patients. Such a dependence would have implications
    for the COVID-19 spreading mechanism, the age-dependent immune system
    reaction, and thus for policymaking. We hereby develop a method to
    analyze viral-load distribution data as a function of the patients'
    age within a flexible, non-parametric, hierarchical, Bayesian, and
    causal model. The causal nature of the developed reconstruction additionally allows
    to test for bias in the data. 
    This could be due to, e.g., bias in patient-testing and data collection or systematic
    errors in the measurement of the viral load. 
    We perform these tests by calculating the Bayesian evidence for each 
    implied possible causal direction. 
    The possibility of testing for bias in data collection and identifying causal directions
     can be very useful in other contexts as well.
    For this reason we make our model freely available.
    When applied to publicly available age and
    SARS-CoV-2 viral load data, we find a statistically significant increase
    in the viral load with age, but only for one of the two analyzed datasets.
    If we consider this dataset, and based on the current understanding
    of viral load's impact on patients' infectivity, we expect a non-negligible
    difference in the infectivity of different age groups. This difference
    is nonetheless too small to justify considering any age group as noninfectious.

% Please keep the Author Summary between 150 and 200 words
% Use first person. PLOS ONE authors please skip this step. 
% Author Summary not valid for PLOS ONE submissions.   
% \section*{Author summary}
% Lorem ipsum dolor sit amet, consectetur adipiscing elit. Curabitur eget porta erat. Morbi consectetur est vel gravida pretium. Suspendisse ut dui eu ante cursus gravida non sed sem. Nullam sapien tellus, commodo id velit id, eleifend volutpat quam. Phasellus mauris velit, dapibus finibus elementum vel, pulvinar non tellus. Nunc pellentesque pretium diam, quis maximus dolor faucibus id. Nunc convallis sodales ante, ut ullamcorper est egestas vitae. Nam sit amet enim ultrices, ultrices elit pulvinar, volutpat risus.       

%\linenumbers

% Use "Eq" instead of "Equation" for equation citations.
\section*{Introduction}
Children do not seem to be major drivers in the transmission of Severe
Acute Respiratory Syndrome Coronavirus 2 (SARS-CoV-2) in the general
population \cite{colson2020children}. However, the exact degree to
which children and adolescents get infected by, and are able to transmit
the virus is not yet well known. Their role
in the community spread depends on their susceptibility, symptoms,
viral load, social contact patterns, behavior, and existing mitigation
strategies as schools and daycares closings. Among all these variables,
the viral load plays a fundamental role. The viral load might help
to predict disease severity \cite{zheng2020viral} and mortality \cite{mortality1,mortality2,mortality3}
and can serve as a proxy for the infectivity of the patient \cite{yonker2020pediatric,jefferson,infectivity}.
The severity of a disease, its infectivity, and its mortality are
certainly fundamental parameters that must be considered when deciding
on best-practice preventative measures to fight the pandemic spread.
Research in this direction can enable truly data-driven policymaking
like, for example, school openings and focused lockdowns. For this
scope, it is important to understand how the viral load depends on
the patients' age.

In this work, we examine viral load as a proxy for infectivity. We
reanalyze the age-stratified viral load data from Jones et al. \cite{jones2020analysis}
in order to better understand the actual relation between these variables.
We do this in the hope of gaining insight into fundamental differences
reported in the literature regarding the relationship between viral
load and age. We achieve this goal by developing a flexible, non-parametric,
causal, and Bayesian model to reconstruct the conditional probability
density function (PDF) of the viral load given the patient's age.
The developed method is a second central result of this work: it can
be applied to future studies on SARS-CoV-2, to similar data from other
diseases, and also to many causally connected quantities in very different
contexts.

The non-parametric PDF reconstruction is regularized by mild assumptions
on the smoothness of the underlying statistical processes. In particular,
we assume that the log-densities are Gaussian-process realizations,
drawn with an a priori flexible correlation kernel parametrized by
a Matérn family correlation function. The parameters of this correlation
function are then inferred along with the PDF through a variational
inference algorithm. To achieve this, we adapt methods developed for
information field theory, the information theory for fields \cite{ensslin09,ensslin18}.
In this context, fields are understood as spatially varying (physical)
quantities. The reconstructed PDFs are regarded as scalar fields whose
values are defined at each point of the two-dimensional space spanned
by age and viral load and represent the probability of observing a
given combination of age and viral load. The toolkit of information field theory
has proven itself to be successful in a wide range of applications,
ranging from 3D tomography \cite{leike2020resolving}, over time-resolved
astronomical imaging \cite{arras2020variable}, to causality inference
\cite{kurthen2020bayesian}.

\paragraph{Outline}

The rest of this work is structured as follows: in Sec.~\nameref{sec:related_work},
we discuss the state of the art of the research on the assessment
of the viral load-vs-age dependence. In Sec.~\nameref{sec:Model-design},
we motivate the need for a causal description and show how this description 
is built into the model and the inference scheme we adopt. 
In Sec.~\nameref{sec:Data}, we describe how the data has been 
acquired and processed. We then outline the main results of our study 
in Sec.~\nameref{sec:Results}, focusing on their impact on the infectivity 
of SARS-CoV-2. 
Finally, in Sec.~\nameref{sec:Conclusions} we summarize the benefits of our 
approach while highlighting its potential limitations and identifying possible 
future work directions.

\section*{Related work\label{sec:related_work}}

Since the outbreak of SARS-CoV-2, efforts have been made in order
to understand whether certain age groups are more susceptible than
others. This could either mean that people from such age groups are
more likely to get infected or that they show more severe symptoms
compared to older or younger individuals. In addition to this, patients
from specific age groups could be more infectious than others, hence
more likely vehiculating the disease. To shed light on these problems,
viral load -- which is a proxy for infectivity -- can be a useful tool.
It is measured by reverse transcription PCR (RT-qPCR) assays from nasopharyngeal 
and oropharyngeal swabs via the so-called cycle threshold (Ct) value. 
The viral load is the virus concentration in the upper respiratory tract
and it is usually expressed as the number of viral RNA copies per mL of sample 
or entire swab specimen or simply by the Ct value. 

Several works have analyzed whether viral loads differ between children and adults 
\cite{10.3389.fmed.2021.608215,jacot2020viral,colson2020children,kleiboeker2020sars,
Euser2021.01.15.21249691,costa2021upper,heald2020age,chung2021comparison,
baggio2020sars,polese2021children,ade2021analysis,cendejas2021lower,
bullard2021infectivity,bellon2021severe}, and between young children 
and adolescents \cite{maltezou2020children,l2020culture,zachariah2020symptomatic},
and have led to conflicting results. At least eight studies from different countries 
have concluded that SARS-CoV-2 viral RNA loads among children and adults were comparable 
\cite{10.3389.fmed.2021.608215,jacot2020viral,colson2020children,costa2021upper,chung2021comparison,
baggio2020sars,polese2021children,ade2021analysis}.
In these studies, dependence between age and viral load has been tested
using one and two-way analysis of variance (ANOVA) \cite{10.3389.fmed.2021.608215}
and median of the viral load. A further study \cite{kleiboeker2020sars}
found that mean and median viral load values did not vary conspicuously
by age but noted that the highest values were measured in patients
born from $1995$ to $2009$. In contrast, at least five studies reported
significant differences in viral loads of young children and adults
\cite{Euser2021.01.15.21249691,costa2021upper,heald2020age,
cendejas2021lower,bullard2021infectivity,bellon2021severe}. 
Euser et al. \cite{Euser2021.01.15.21249691}
suggested an approximately $16$-fold higher viral load in the oldest
age group ($>79$ years), compared to the youngest age group ($<12$
years). Here, age-group differences in the viral load distribution
are assessed making use of the Kruskal-Wallis test and linear regression.
Another work \cite{heald2020age} estimated the amount of SARS-CoV-2
in the upper respiratory tract of young children ($<5$ years) to
be $10$-fold to $100$-fold greater than in adults whereas a work
from Zachariah et al. \cite{zachariah2020symptomatic} showed that mean viral load
was significantly higher in infants ($<1$ year) as compared to older
children and adolescents. 

Finally, one of the largest and most widely
followed studies on the subject -- even though the number of children
and adolescents included is fairly small - was carried out by Jones et al. \cite{jones2020analysis}
in early 2020. Dependence between viral load and age has been tested
for different age groups both as categorical data and treating age
as a continuous variable. In order to compare the viral load of different
age categories, the categorical data has been analyzed in a parametric
(Welch\textquoteright s T-test), non-parametric (Mann-Whitney rank
test), and Bayesian fashion (modeling viral loads as a mixture of
gamma distributions). When considering age as a continuous variable,
viral loads have been predicted from age, type of PCR system, and
age-PCR system interaction. This study from Jones et al. \cite{jones2020analysis}
did not reveal large differences in the viral loads of different age
groups, a result that was publicly debated in Germany for its possible
implications for school opening policies. In the following, we reanalyze
this data. In a very recent publication, Jones et al. \cite{Joneseabi5273} extended 
their initial version of the study. In this newer version, they make use 
of thin-plate spline regression to conclude that children and adolescents 
have a slightly lower viral load than adults, but that this difference is unlikely to be clinically relevant.

None of the studies in the literature utilizes the causal framework.
In this work, we choose to adopt this framework since it naturally allows to answer the
central question of whether the age of a patient causes its viral load and infectivity.
It also leads to additional advantages that we will present in the following sections.
The problem of deducing causal directions (in particular for the bivariate 
$x \rightarrow y$, $y \rightarrow x$ and $x \perp y$ case) from observational data
coming from a joint distribution has been introduced by the works of J. Pearl \cite{reason:Pearl09a}
and P. Spirtes et al. \cite{RePEc:mtp:titles:0262194406}, further developed by 
Mooij, J. et al. \cite{JMLR:v17:14-518} and is now a central and 
non-trivial problem in data analysis.
For additional details and motivations behind causal inference
theory and techniques we refer to their works.

\section*{Model design\label{sec:Model-design}}

In Sec.~\nameref{sec:related_work} we described the state of the art
of statistical analyses performed in order to investigate the age
dependence of the viral load. These analyses mostly rely on variance
tests or correlation assessments. In the following, we motivate the
need for a causal model of the viral load distribution. In fact, in
order to explain the age and viral load data collected by Jones et al. \cite{jones2020analysis},
a well-behaved model should incorporate basic knowledge about the
causal relation between age and viral load. It should furthermore
allow questioning whether the viral load distribution depends on the
patients' age and quantify the strength of such dependence, if it
exists. We consequently develop a non-parametric causal model and
apply it to data.

\subsection*{Motivation}
In order to make statements about the factors that contribute
to the spread of a disease, we need to rely on data to ground these claims.
More importantly, we need to find a model that is capable of identifying 
and explaining the relationships that underlie the data. 
Indeed, this should be a minimal requirement for any data-analysis task. 
The choice of an incorrect model can lead to wrong or misleading conclusions.
This is clearly an issue regardless the nature of the data at hand.
However, in the case of data describing the pandemic spread the choice of
an incorrect model can lead to ineffective or even potentially harmful decisions. 
It is therefore of vital importance to be able to capture - at least up to
a certain level of uncertainty - the interdependences underlying the data.

As discussed in the previous section Sec.~\nameref{sec:related_work}, 
the problem of identifying the relation between age and viral load has been 
tackled making use of many different techniques. 
While categorical data analysis and linear correlation analysis struggle to 
pick up all possible dependences, non-parametric approaches have the 
drawback of possibly being too (or too little) complicated to account for the actual functional
dependence that is inherent to the data.
For this reason we want to build a model that is flexible in the sense
that it can simply reconstruct independent densities with few degrees of freedom
while also being capable of inferring more complex distributions, 
when needed to describe the data.
Moreover, we would like the model to automatically be able to
adjust its complexity. This way, the choice of the model is independent
from the data analyst's choices and the results are more consistent 
and reproducible.

This is another benefit of adopting the causal framework.
The concepts of ``simple" independence and ``more complicated" dependence 
between variables are natural to this framework and are described by
the causal graph and the structural causal model of choice \cite{JMLR:v17:14-518}.
Furthermore, claiming causal dependence is much stronger than simple 
linear correlation, which also can arise from a confounder or by mere coincidence.
Causal dependence is directional. This is a particularly interesting
feature as it allows to test for bias in the data-collection process when a
unnatural causal direction is detected from the data analysis, as we will discuss in
Sec.~\nameref{sec:Results}.

\subsection*{Causal structure\label{subsec:Causal-structure}}

\begin{figure*}[t]
\centering
\includegraphics[width=0.9\textwidth]{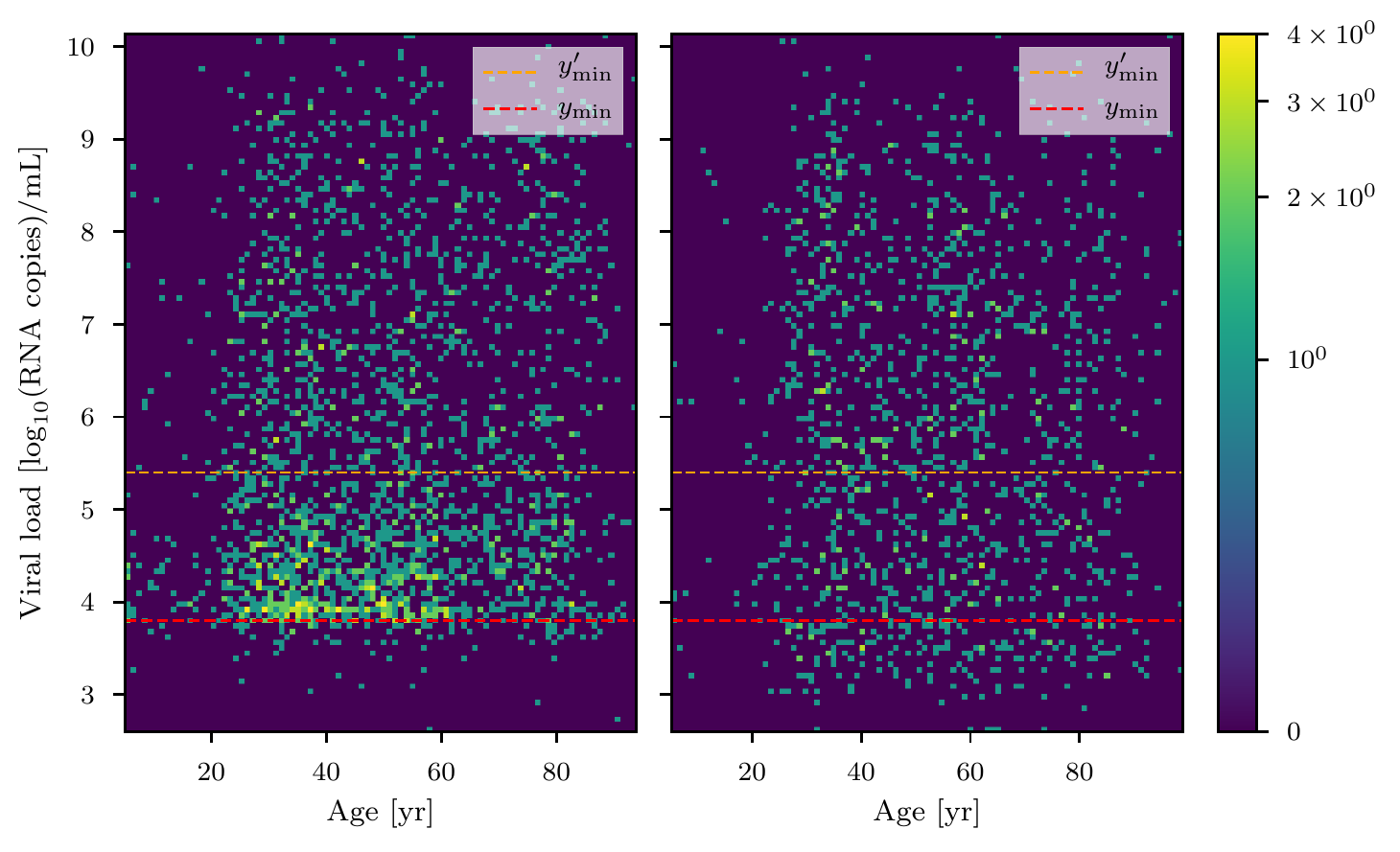}

\caption{\textbf{The cobas (left) and the LC 480 (right) datasets}. The lower thresholds
$y_{\text{min}}$ and $y'_{\text{min}}$ with which the data has bene
filtered are shown in red and orange, respectively. The number of data counts is color coded with
a logarithmic color scheme (see colorbar).}
\label{fig:data-thresholds}
\end{figure*}

\begin{figure}[ht]
    \centering
    \includegraphics[width=0.8\textwidth]{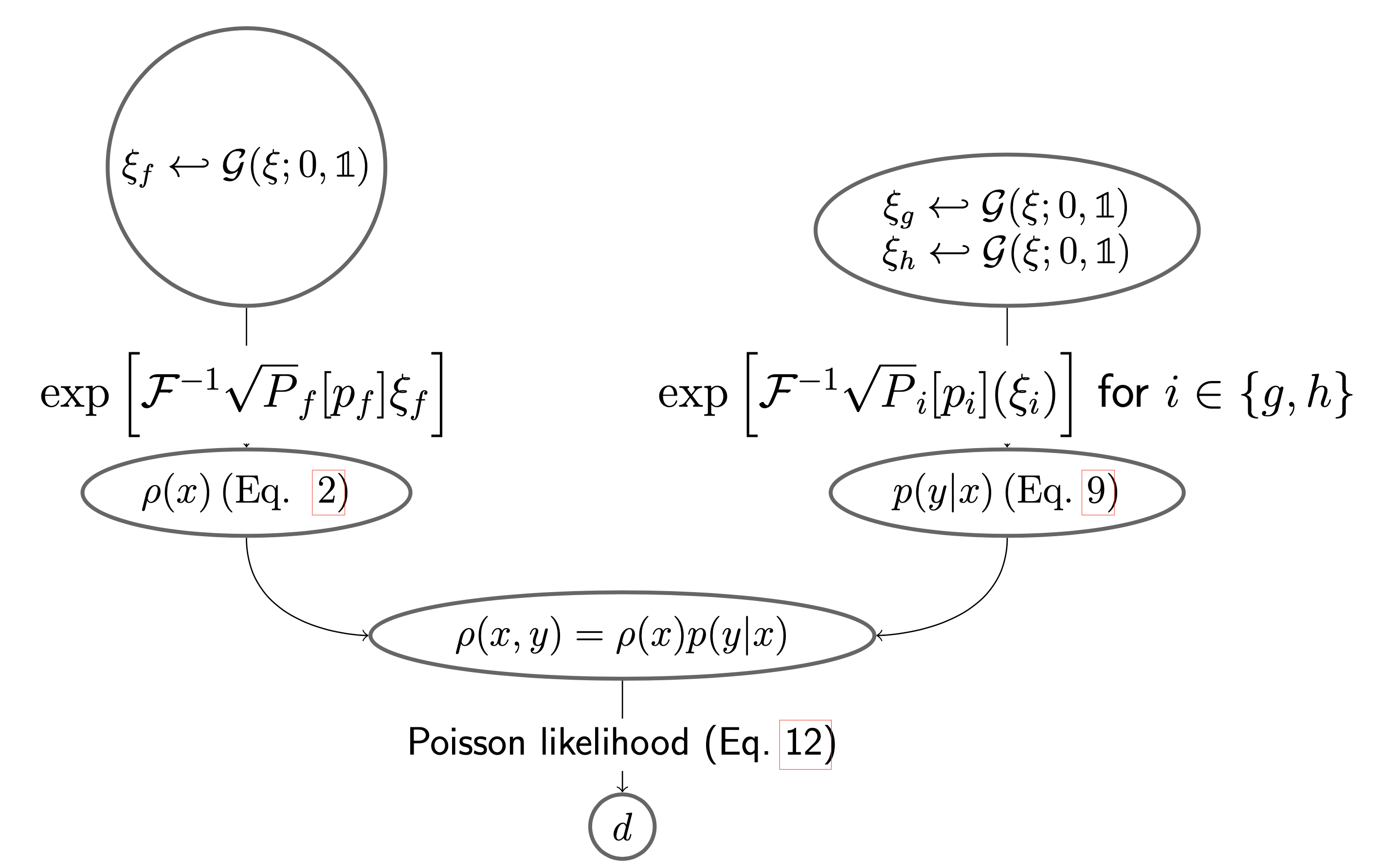}
    \caption{\textbf{Graph structure of the causal model of age $\vb{x}$ and viral load $\vb{y}$}. $\mathcal{F}$ denotes the Fourier transform operator. 
    Starting from the top, standard-normally distributed excitations $\xi$ are drawn from the latent priors for $f, g,$ and $h$. They are then transformed
    into the non-parametric signal $\rho(x,y)$ (Eq.~\ref{eq:pdf_cond}) by taking the inverse Fourier transform of the Matérn kernel parametrization 
    described in Eq.~\ref{eq:power-spec}. Finally, the signal $\rho(x,y)$ is compared to the data $d$ through a Poissonian likelihood (Eq.~\ref{eq:poisson_like}).}
\label{fig:causal_graph}
\end{figure}

The analyzed dataset $d=\{(i,x_{i},y_{i})\}_{i=1}^{N}$ consists of
indexed pairs of age $x_{i}=\text{age}_{i}/\text{years}$ and log-viral
load $y_{i}=\log_{10}(\text{viral RNA copies}_{i}/\text{ml)}$ values
for each of the $N$ infected patients, extracted from Fig.\ 6 in
\cite{jones2020analysis}. The dataset is shown in Fig.~\ref{fig:data-thresholds}.
For simplicity, we refer to $y$ as the ``viral load'', where $10^{y}$
are the number of viral RNA copies per milliliter of sample or entire
swab specimen. We assume that the data points are Poisson process
counts drawn from an underlying stationary density distribution $\varrho(x,y)$.
Using the causal model shown in Fig.~\ref{fig:causal_graph} we reconstruct the density $\varrho(x,y)$. 
Our final density reconstructions are displayed in Fig.~\ref{fig:data-reconstruction} and Fig.~\ref{fig:data-reconstruction-lc}.
To model this density, we express it in terms of the underlying age 
distribution $\varrho(x)$ of the patients times the conditional 
PDF $p(y|x)$ of the viral load $y$ given the patient's age $x$,
\begin{equation}
    \varrho(x,y)=\varrho(x)\ p(y|x).\label{eq:split}
\end{equation}
In Eq.~\ref{eq:split} the causal direction $x\rightarrow y$ (age
influences viral load) is implicitly introduced. Even though $x\rightarrow y$
might appear to be the most intuitive causal direction, since the
immune system reaction depends on age and thus age should affect the
viral load, we note that selection effects could introduce different
apparent causal structures in the data. If
the data had been collected in such a way that the viral load
($y$) was the deciding factor for whether a patient would enter the
data sample, with an age ($x$) dependent threshold, the viral load
would impact the age distribution in the sample, leading to an apparent
$y\rightarrow x$ causal structure. 

As a practical example, one could imagine physicians to request a PCR test 
for a child only if their symptoms were worse compared to those of an adult
for which they would have requested a PCR test. 
Since this and similar selection effects cannot be fully excluded 
for the analyzed data (see discussion in \cite{jones2020analysis}),
we will calculate the Bayesian evidence for the possible causal relations
$x\rightarrow y$, $y\rightarrow x$, and $x\perp y$ (i.e. 
$x$ and $y$ are independent, $p(y|x)=p(y)$).
As a final remark, we note that other effects such as, e.g., delays in 
data reporting, could in principle also be confounding the simple $x\rightleftarrows y$
causal direction. Li and White~\cite{10.1371/journal.pcbi.1009210} have
for instance shown how reporting delays can have an effect on the pandemic-spread models.
Effectively, a confounding variable $z$ would influence both 
$x$ and $y$, namely $x \leftarrow z \rightarrow y$.
Although we cannot completely exclude the presence of such a confounding
variable, the randomization tests that we describe in Subsec.~\nameref{subsec:causal_results}
support our implicit assumption of absence of any leading confounding variable.

To model the $x\rightarrow y$ causal direction, we need to model
the age distribution $\varrho(x)$ of the infected patients according
to the causal structure introduced in Eq.~\ref{eq:split}. Lacking
knowledge on the exact details of the patient selection process, we
assume $\varrho(x)$ to be a log-normally distributed random variable.
The log-normal distribution is a natural choice since an age density
is by definition a strictly positive and continuous quantity. Another
natural assumption is the absence of abrupt changes, since no sharp
age-selecting processes are expected to have shaped it. We fulfill
these assumptions with the choice 
\begin{equation}
    \varrho(x)=\varrho_{0}e^{f(x)},\label{eq:x-marginal-1}
\end{equation}
where ${\varrho_{0}=N/100}$ is a reference density and ${f:[0,100]\mapsto\mathbb{R}}$
a smooth function centered around zero. We accordingly assume $f$
to be drawn from a zero centered Gaussian process with a prior covariance
$F$ 
\begin{equation}
\mathcal{P}(f)=\mathcal{G}(f,F):=\frac{1}{\sqrt{2\pi F}}\exp\left(-\frac{1}{2}f^{\dagger}F^{-1}f\right).\label{eq:Gauss}
\end{equation}
The covariance 
\begin{equation}
    F_{xx'}= \expval{f(x)\,f(x')}_{(f)}:= \int\mathcal{D}f\ \mathcal{P}(f)\,f(x)\,f(x')
\end{equation}
determines the degree of smoothness of the logarithmic distribution
function, as well as the characteristic length scale and the amplitude
of its variations. We assume this correlation structure to be translation
invariant $F_{xx'}=F(x-x')$, since we only expect it to depend on
age differences - and not on a particular age value - and parametrize
it with a Matérn kernel. Invoking the Wiener-Khinchin theorem, we
can represent such a translation-invariant correlation function in
Fourier space with a spectral density of 
\begin{equation}
    P_{f}(k)=\frac{a_{f}^{2}}{\left[1+(k/k_{f})^{2}\right]^{\nicefrac{\gamma_{f}}{2}}},
    \label{eq:power-spec}
\end{equation}
with $a_{f}$ specifying the amplitude of the variations in $f$,
$1/k_{f}$ the characteristic length-scale above which the variations
become uncorrelated, and $\gamma_{f}$ the spectral index, which determines
the smoothness of the variations. We infer all three covariance parameters
$p_{f}\coloneqq(a_{f},k_{f},\gamma_{f})$ from the data. In order
to ensure that the model is flexible enough to fit the data, we set
mildly informative priors on the covariance parameters.
We denote by $\mathcal{P}(f|p_{f})$ the probability of a specific
realization of $f$ given the Matérn kernel parameters $p_{f}$, as
described by Eqs.\ \ref{eq:Gauss} to \ref{eq:power-spec}.

Next, we have to specify the distribution of the viral load given
the age, $p(y|x)$. To do so, we note that we can directly model 
an independent distribution for which $p(y|x)=p(y)$ in the same
way as we modeled $\rho(x)$, i.e. by choosing $p(y|x) \propto e^{g(y)}$.
Of course, $p(y|x)$ can in general exhibit an arbitrary complicated
dependence on  $x$. We model any additional complicated 
entanglement between the age $x$ and the viral load $y$ with 
a new function $h(x,y)$. By doing so, we introduce a degeneracy 
between $g(y)$ and $h(x,y)$, since they can both model $y$-only
dependent structures. In principle, the function $h(x,y)$ can in fact model 
any function $p(y|x)$ without the necessity of introducing $g(y)$. 
To solve this problem, we choose
\begin{equation}
    p(y|x)\propto\frac{e^{g(y)+h(x,y)}}{\int e^{h(\tilde{x},y)}\dd{\tilde{x}}}.\label{eq:P(y|x)}
\end{equation}
To ensure that only $g(y)$ models strictly $y$-dependent features and 
that all possibly complicated $x-$dependent features are captured by $h(x,y)$, 
we have to prevent $h(x,y)$ from modeling any strictly $y$-dependent structure 
that has been already captured by $g(y)$. 
We do this with the denominator $\int e^{h(\tilde{x},y)}\dd{\tilde{x}}$ in Eq.\ \ref{eq:P(y|x)},
that removes from $p(y|x)$ any $y$-only structure contained in $h(x,y)$, 
hence eliminating the degeneracy between $g(y)$ and $h(x,y)$.

We can verify that $h(x,y)$ indeed satisfies the desired property 
of not encoding any structure that could be represented 
by $g(y)$ by simply substituting 
$h(x,y)\mapsto h'(x,y)\coloneqq h(x,y)+g'(y)$ in Eq.\ \ref{eq:P(y|x)}, 
where $g'(y)$ depends only on $y$. This results in $h$ and $h'$ 
leading to the same conditional PDF 
\begin{align}
\begin{split}
    p'(y|x) &\propto\frac{e^{g(y)+h'(x,y)}}{\int e^{h'(\tilde{x},y)}\dd
    {\tilde{x}}}=\frac{e^{g(y)+h(x,y)+\cancel{g'(y)}}}{\cancel{e^{g'(y)}}\int e^{h(\tilde{x},y)}\dd{\tilde{x}}}\\
    &=\frac{e^{g(y)+h(x,y)}}{\int e^{h(\tilde{x},y)}\dd{\tilde{x}}}\propto p(y|x)    
\end{split}
\label{eq:cond_prob}
\end{align}
and we can therefore conclude that only $g(y)$ can model 
strictly $y$-dependent features.

As desired, for vanishing $h$ it still holds 
\begin{eqnarray}
    \eval{p(y|x)}_{h(x,y)=0} & \propto & e^{g(y)} \propto p(y),
\end{eqnarray}
which implies independence, $y\perp x$. Thus, a non-trivial $h(x,y)$ 
models the $x\rightarrow y$ causal influence while a trivial $h(x,y)\equiv0$
represents causal independence.
We now want to make sure that the more complicated $x-$dependence
modeled by $h(x,y)$ is only introduced if it is strictly needed to explain the data. 
This way, we can clearly distinguish the causal scenarios $x\rightarrow y$ 
or $y\rightarrow x$ from the independent $x\perp y$ scenario.
In fact, the distinction between $g$ and $h$ would be meaningless without
a prior choice that favors independence between $x$ and $y$. 
Hence, we assume $g$ and $h$ to be drawn from zero-centered 
Gaussian processes.
In this way, without any information coming from the data, the most
likely realizations of both of these functions are 
the identically-zero functions $g(y)=h(x,y)\equiv0$.
However, if the data exhibits strictly $y$-dependent features, 
meaning that the marginal distribution $p(y)$ is non-trivial, 
these features can only be represented by a non-zero $g(y)$. 

Strictly $y$-dependent features are indeed clearly visible
in the data. For example, we notice that in Fig.\ \ref{fig:data-reconstruction}
higher viral loads are by far more rare than lower viral loads.
In this case, the data-inferred $g(y)$ is non-zero and shows this decreasing feature.
Following the same reasoning, the most likely distribution for $h(x,y)$
in absence of data is identically-zero everywhere. Again, $h(x,y)$
will only be non-zero in case that the data triggers some coupling
between $x$ and $y$. Thus, the model favors independence of $x$
and $y$ (by favoring $h\equiv0$) and the inferred density will be
entangled in $x$ and $y$ only in the case the data exhibits this
feature. This shows how the level of complexity of the
model is adapted to the data automatically, without having to make any
additional explicit model choice.%\par\medskip
 
We again assume a Matérn-kernel-shaped correlation structure for the
Gaussian process $g$, with covariance parameters ${p_{g}\coloneqq(a_{g},k_{g},\gamma_{g})}$.
We set the priors on $p_{g}$ as for $p_{f}$ and learn these parameters
from the data as well. Since the typical length scales and amplitudes
of the variations of $h(x,y)$ in $x$ and $y$ directions are not
in principle a priori similar, we assume the covariance for $h$ to
be shaped by a direct product of individual Matérn kernels in the
$x$ and $y$ directions. For their corresponding prior parameters,
$p_{h}=(p_{h}^{(x)},p_{h}^{(y)})$ with $p_{h}^{(i)}=(a_{h}^{(i)},k_{h}^{(i)},\gamma_{h}^{(i)})$
and $i\in\{x,y\}$, we use similar hyper-priors as before, i.e. as
for $p_{f}$ and $p_{g}$, respectively. For more details on the prior choices
we refer to the Appendix \nameref{sec:Mat=0000E9rn-kernel-density}.
We call the ensemble of all
these kernel parameters $p\coloneqq(p_{f},p_{g},p_{h}).$ The details
on how the Gaussian process for $h$ with a Matérn kernel product
covariance structure is set up is described in the Appendix \nameref{sec:Mat=0000E9rn-kernel-density}, where we describe a multi-dimensional 
density estimator that is agnostic to causal directions.

Lastly, we normalize the conditional PDF 
\begin{equation}
    p(y|x)=\frac{e^{g(y)+h(x,y)}}{\int e^{h(\tilde{x},y)}\dd{\tilde{x}}}\ \left(\int\frac{e^{g(\tilde{y})+h(x,\tilde{y})}}
    {\int e^{h(\tilde{x},\tilde{y})}\dd{\tilde{x}}}\dd{\tilde{y}}\right)^{-1}
    \label{eq:pdf_cond}
\end{equation}
such that the full model density reads 
\begin{equation}
    \varrho(x,y)=\varrho_{0}e^{f(x)}\ \frac{e^{g(y)+h(x,y)}}{\int e^{h(\tilde{x},y)}\dd{\tilde{x}}}\ 
    \left(\int\frac{e^{g(\tilde{y})+h(x,\tilde{y})}}{\int e^{h(\tilde{x},\tilde{y})}\dd{\tilde{x}}}\dd{\tilde{y}}\right)^{-1}.
\end{equation}
A schematic representation of the forward causal model is shown in Fig.~\ref{fig:causal_graph}.
The assumed causal structure ${x\rightarrow y}$ is introduced in
the model by the asymmetry between the roles of the $x$ and $y$
coordinates and the zero-centered Gaussian process priors on $f$,
$g$, and $h$. Interchanging $x$ and $y$ leads to a model that
follows the opposite causal direction ${y\rightarrow x}$. This allows
to empirically distinguish these causal directions by calculating
the model evidences for the two opposite scenarios, namely ${x\rightarrow y}$
and ${y\rightarrow x}$, as well as to test for $x\perp y$ by enforcing
$h=0$. 

The evidence already inherenty penalizes more complicated
models (see~\cite{andrijaspaper}), i.e. models that require a higher 
number of degrees of freedom. 
Using the ELBO as a criterion for model selection has therefore the additional 
benefit of regularizing the solutions.
In other words, the evidence further helps the data analyst to univocally pick the lowest 
complexity model which can best explain the data, in addition to the already self-adjusting level 
of complexity provided within the model itself. 

Finally, we note that the choice of $x$ and $y$ is up to now completely
arbitrary. Therefore, the method described in this work can be used to 
assess the causal direction of any continuous two-dimensional data distributions, 
given the data and some loose (prior) information about the correlation structure. 
the proposed method is thus very general and  can be used for different datasets 
and analysis schemes with respect to those described in this paper. 
For example, it could be used to test the causal relation between the concentration 
of greenhouse gasses in the atmosphere and the average global temperature rise
in Celsius degrees.

\subsection*{Likelihood}

In order to construct the likelihood $\mathcal{P}(d|\varrho(\cdot,\cdot))$,
we bin the data into a fine two dimensional grid over the $x$ and
$y$ coordinates with $90\times128$ pixels, such that 
\begin{align*}
n_{ij}(d)=\sum_{m=1}^{N}\int_{i\Delta x}^{(i+1)\Delta x} \dd{x} \int_{j\Delta y}^{(j+1)\Delta y}\delta(x-x_{m})\delta(y-y_{m}) \dd{y}
\end{align*}
contains the number of cases within the $(i,j)^{\text{th}}$ pixel
of size $\Delta x=1\,\text{yr}$ and $\Delta y\simeq0.04\log_{10}(\text{viral RNA copies}_{i})/\text{ml}$.
These counts $n_{ij}$ are then compared with the model's expectations
\begin{align}
\begin{split}
    \lambda_{ij} &\coloneqq \lambda_{ij}(\varrho)=\int_{i\Delta x}^{(i+1)\Delta x}\dd{x}\int_{j\Delta y}^{(j+1)\Delta y}\varrho(x,y) \dd{y}\\
 & \approx\Delta x\Delta y\ \varrho\qty(\qty(i+\frac{1}{2})\Delta x, \qty(j+\frac{1}{2})\Delta y)
\end{split}
\end{align}
via a Poisson likelihood 
\begin{equation}
    \mathcal{P}(d|\varrho)=\prod_{i,j}\frac{\lambda_{ij}^{n_{ij}}}{n_{ij}!}e^{-\lambda_{ij}}.
    \label{eq:poisson_like}
\end{equation}

\subsection*{Inference\label{subsec:Inference}}

The full model involving the data $d$ as well as all the unknown
quantities, which compose the signal vector ${s\coloneqq(f,g,h,p)},$
reads 
\begin{align}
\begin{split}
    \mathcal{P}(d,s) &= \mathcal{P}(d|s)\,\mathcal{P}(s), \quad\text{where}\\
    \mathcal{P}(d|s) &= \mathcal{P}(d|\varrho[f,g,h]) \quad\text{and}\\
    \mathcal{P}(s) &= \mathcal{P}(f|p_{f})\mathcal{P}(g|p_{g})\mathcal{P}(h|p_{h})\,
    \mathcal{P}(p_{f})\,\mathcal{P}(p_{g})\,\mathcal{P}(p_{h}). 
\end{split}
\end{align}
At this stage, we need to convert our causal model into an inference
machine for the signal vector $s$. We do this conversion by reformulating
the model in the language of information field theory \cite{ensslin09,ensslin18},
transforming the coordinates of the signal vector $s=s(\xi)$ such
that the prior on the new $\xi$ coordinates becomes an uncorrelated
Gaussian $\mathcal{P}(\xi)=\mathcal{G}(\xi,\mathbb{1})$ as described
by Knollmüller and Enßlin \cite{2018arXiv181204403K}. We then implement the resulting
model using the Python package Numerical Information Field
Theory (\texttt{NIFTy})\cite{missing1, missing2, missing3}
and finally use \texttt{NIFTy}'s implementation of Metric Gaussian
Variational Inference (MGVI) \cite{2019arXiv190111033K} to approximate
the posterior distribution in the new coordinates 
\begin{equation*}
    \mathcal{P}(\xi|d)=\frac{\mathcal{P}(d|\xi)\mathcal{P}(\xi)}{\mathcal{P}(d)}\approx\mathcal{G}
    (\xi-\overline{\xi}_{d},\Xi_{d})    
\end{equation*}
with a Gaussian which has posterior mean $\overline{\xi}_{d}$ and
covariance $\Xi_{d}$, where the $d$ suffix indicates the dataset
used in the inference. This Gaussian posterior encodes the approximate
result of the inference in the new coordinates. In order to translate
this into the signal coordinates, we have to transform $\mathcal{P}(\xi|d)$
to $\mathcal{P}(s|d)$ using the relation $s=s(\xi)$. This relation
is non-linear and the resulting PDF is neither Gaussian nor practical
to obtain analytically. In order to evaluate moments from the posterior
distributions of the desired quantities, MGVI provides $\xi$-samples
drawn from the approximate Gaussian posterior $\xi\hookleftarrow\mathcal{G}(\xi-\overline{\xi}_{d},\Xi_{d})$.
These $\xi$-samples can be converted via the coordinate transformation
$s=s(\xi)$ into the signal space, where they represent (approximate)
signal posterior samples. Making use of these posterior samples it
is possible to calculate the posterior expectation values and model
uncertainties of any desired quantity $q(s)$: 
\begin{align}
\begin{split}
    \overline{q} &\coloneqq \expval{q(s)}_{(s|d)}\approx\frac{1}{N_{\text{s}}}\sum_{i=1}^{N_{\text{s}}}q(s(\xi_{i}))\\
    \sigma_{q}^{2} & = \expval{(q(s)-\overline{q})^{2}}_{(s|d)}.
\end{split}
    \label{eq:mgvi_sigma}
\end{align}
Here, $\xi_{i}$ denotes the $i^{\text{th}}$ of the $N_{\text{s}}$
drawn samples. In particular, the posterior mean of the conditional
PDF $p(y|x)$ and of any quantity which can be calculated from $p(y|x)$
can thereby be obtained, as well as the resulting uncertainties characterized
via their uncertainty dispersion. In general, these posterior signal
samples will not follow Gaussian statistics because the transformation
is typically non-linear. Furthermore, since MGVI is a variational
inference approach, the calculated uncertainties will be slightly
smaller compared to the ones given from the accurate posterior. However
given the complexity and size of the model, we need to use an approximate
inference method as MGVI. For details about this methodology, as well
as extensive performance and accuracy tests, we refer to Knollmüller and 
Enßlin \cite{2019arXiv190111033K}.
\begin{figure*}[t]
\includegraphics[width=1\textwidth]{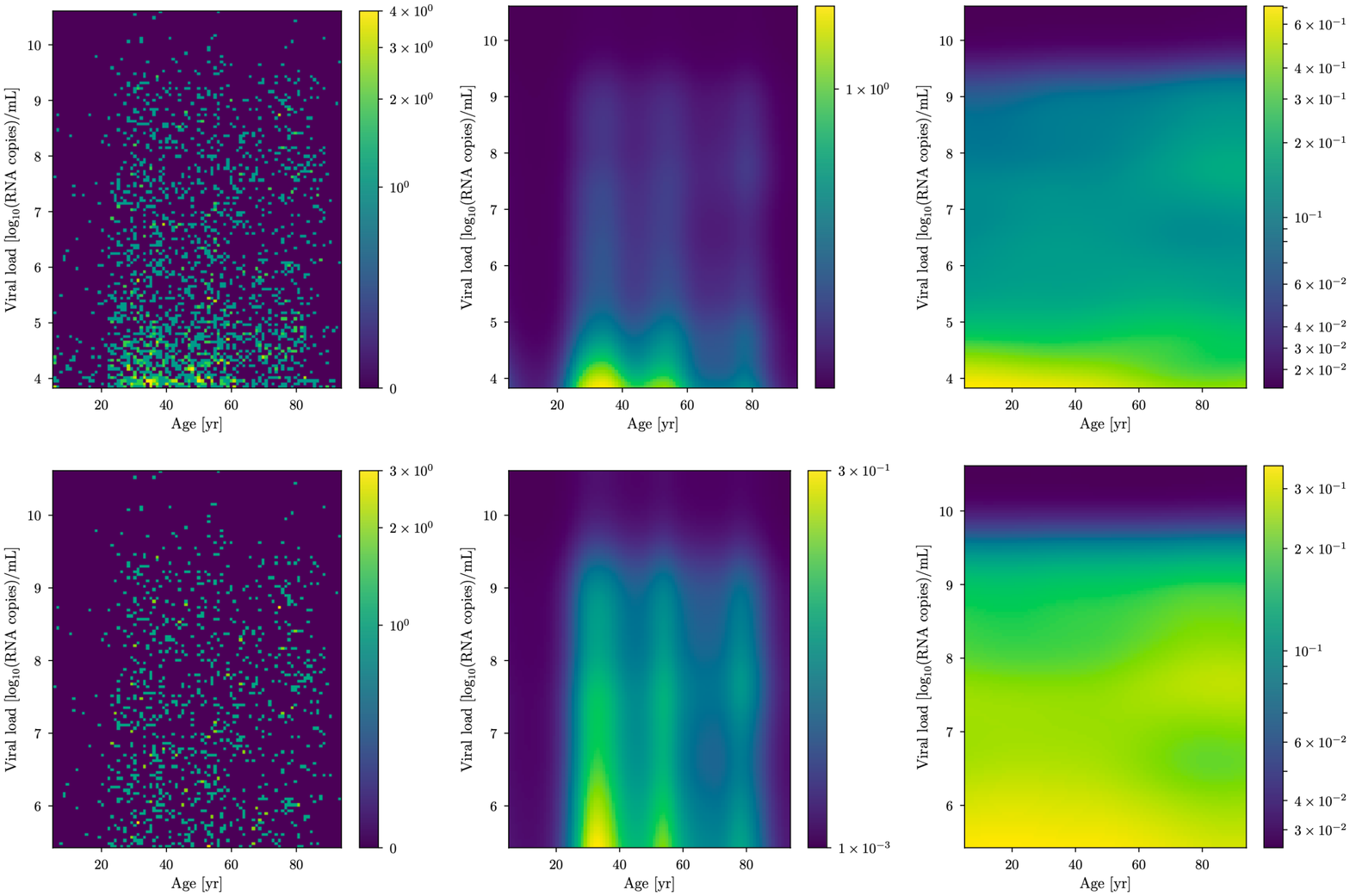}
\caption{\textbf{Cobas dataset analysis.} Left: The cobas dataset. Middle: the 
reconstructed density distribution $\varrho(x,y)$
as a function of the age ($x$) and the viral load ($y$) in a logarithmic
coloring scheme. Right: The $2$D conditional probability distribution
$p(y|x)$ of the viral load (Eq.~\ref{eq:pdf_cond}) obtained by fitting the model.
The data and the results of the analysis are shown for two different 
data-filtering thresholds $y_{\text{{min}}}$ (top) and $y'_{\text{{min}}}$ (bottom).}
\label{fig:data-reconstruction}
\end{figure*}
\begin{figure*}[t]
\includegraphics[width=1\textwidth]{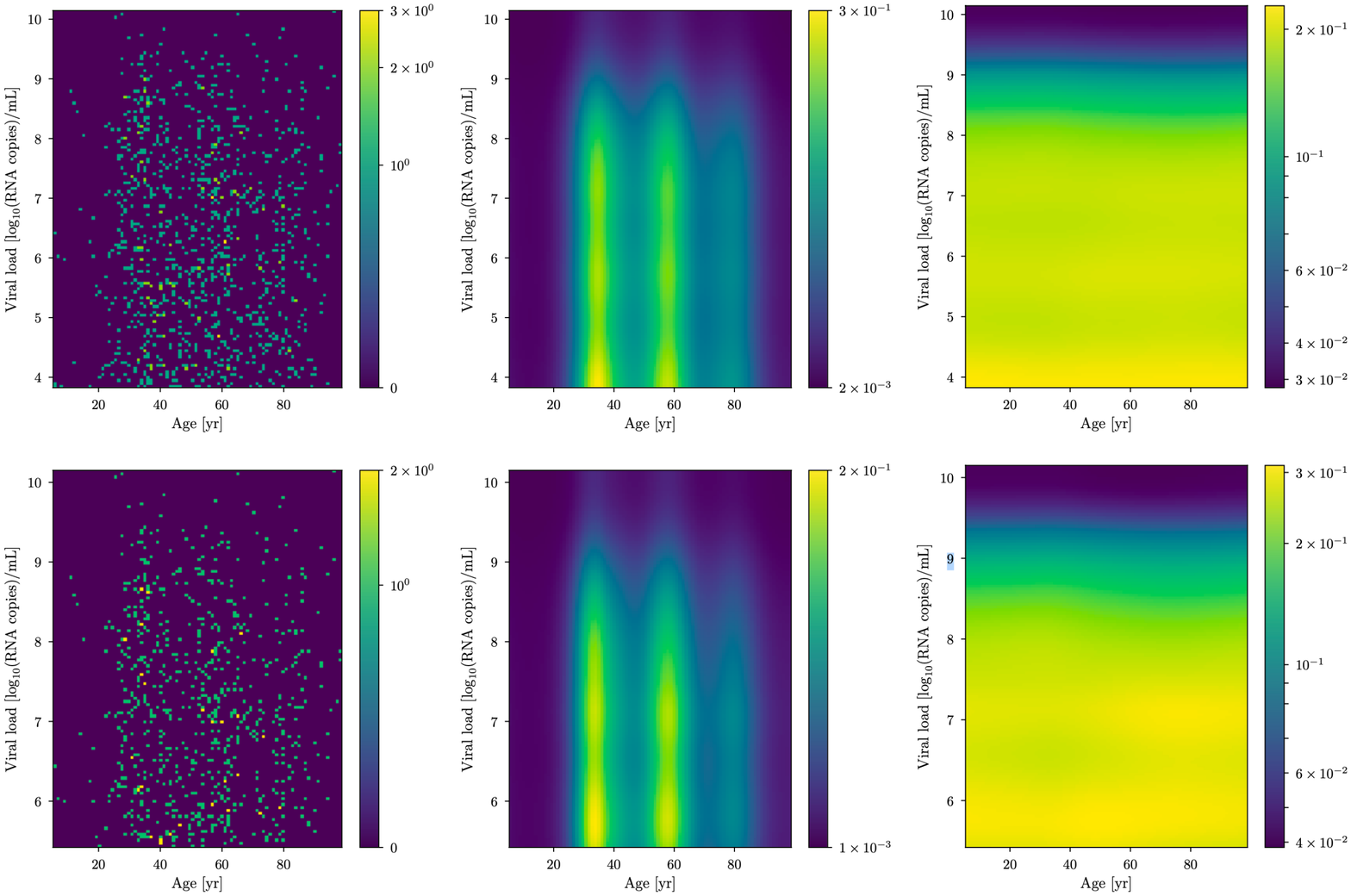}
\caption{\textbf{LC 480 dataset analysis.} The same as in Fig.~\ref{fig:data-reconstruction}, but for the
LC $480$ dataset.}
\label{fig:data-reconstruction-lc}
\end{figure*}

\subsection*{Identifying causal directions with the Evidence Lower Bound \label{subsec:elbo}}

We now focus on understanding the causal relations given by the interplay
between the variables. The proposed causal model should allow to discriminate
between all possible causal structures, namely $x\rightarrow y$,
$y\rightarrow x,$ and $x\perp y$. Since we have a different model for
 each of these causal directions, we can use the Bayesian evidence
to select the model that is better suited to the data.  
We do this once again by exploiting our variational inference scheme.
It has been shown that the evidence is indeed a consistent and robust 
criterion for model selection \cite{knuth, pmlr-v96-cherief-abdellatif19a}.

First, we need to define the independent model, i.e. a model for which age
and viral load are regarded as statistically independent variables.
Such a model is built by setting a very tight zero-centered prior
on $h$, hence removing it from the model for all practical purposes.
We can then estimate the Bayesian evidence both for the causal --
hence dependent -- models ($x\rightarrow y$ and $y\rightarrow x$)
and compare it with the evidence of the independent model ($x\perp y$).
More precisely, we calculate the so called Evidence Lower Bound (ELBO)
as a proxy for an evidence. Using the posterior uncertainty 
covariance $\Xi$ as well as the posterior samples provided by MGVI, 
we can compute the ELBO~\cite{andrijaspaper} for each model. 
This is lower than the exact logarithm of the evidence
by the information difference (as measured in nits) between the exact
and approximated posterior of a model. If too much information
is not lost during MGVI, the ELBO should be a good approximation 
of the exact log evidence. Furthermore, we can assume that deviations from the
exact log evidence should be similar among different models, thereby
reducing the effect of the MGVI approximations on differences between
log-evidences. Thus, we can use the ELBO as a good proxy for the log
model evidence ratios of similar models. The stochastic sampling steps
performed in order to estimate the ELBO introduce a sampling uncertainty.
This uncertainty can be in principle reduced by taking more samples,
at the expense of larger computational costs. We state this numerical
one-sigma uncertainty for all MGVI and ELBO based log-evidences.
The $y\rightarrow x$ model is obtained by swapping the coordinates 
of the $x\rightarrow y$ model. The quantity of interest is then the
logarithm of the evidence ratio of each causal model with the independent
one,
\begin{equation}
    \Delta E_{x\rightleftarrows y}=\log\frac{p(d|x\rightleftarrows y)}{p(d|x\perp y)},
\end{equation}
where $x\rightleftarrows y$ denotes either $x\rightarrow y$ or $y\rightarrow x$.
$\Delta E_{x\rightarrow y}$ indicates the log evidence in favor of
the causal model $x\rightarrow y$ with respect to the independent
one and similarly $\Delta E_{y\rightarrow x}$ the one for $y\rightarrow x$.
Comparing $\Delta E_{x\rightarrow y}$ with $\Delta E_{y\rightarrow x}$
also allows to discriminate between the two possible causal directions
in the dataset. We identify the preferred causal direction underlying the data
 in subsec.~\nameref{subsec:causal_results}, where we also discuss
 its implications.

\section*{Data\label{sec:Data}}

We make use of RT-PCR viral load data collected from the Charité Institute
of Virology and Labor in Berlin, Fig.\ 6 in Jones et al. \cite{jones2020analysis}.
The data was acquired with two different PCR instruments, Roche cobas
$6800$/$8800$ (cobas dataset, which we denote by $d_{\text{C}}$
and is comprised of $\approx2200$ data points) and Roche LightCycler
$480$ II (LC $480$ dataset, which we denote by $d_{\text{L}}$ comprised
of $\approx1350$ data points). In the following, we will show the
difference between the two datasets. As can be seen from the count
difference in the raw data plots as of Fig.\ 6 of \cite{jones2020analysis}
and in the histogramed data in Figs.~\ref{fig:data-reconstruction}
and \ref{fig:data-reconstruction-lc}, for low viral loads ($y\lesssim5$
in units of $\log_{10}(\text{RNA copies/ml})$), the LC $480$ dataset
shows a roughly uniform count distribution in the whole viral load
domain. In contrast, the cobas dataset exhibits an increasing number
of counts in the $y\in[2.0,3.8]$ viral load domain followed by a
descending trend in counts in the $y\in[3.8,5.0]$ region. For this
reason and in order to better understand the possible shortcomings
of both instruments, we analyzed the data in two different ways. 

Since the major differences between the datasets arise for viral load values
$y\in[3.8,5.0]$ we define two lower thresholds for the viral load
($y_{\text{min}}\coloneqq3.8$ and $y'_{\text{min}}\coloneqq5.4\simeq\log_{10}(250000)$
in units of $\log_{10}(\text{viral RNA copies/ml})$) and discard
any data point for which the viral load is lower than $y_{\text{min}}$
and $y'_{\text{min}}$, respectively. We set the lower threshold $y_{\text{{min}}}$
at the value for which the number of counts of the cobas dataset is
maximum (see Fig.~\ref{fig:data-thresholds}). Below this threshold,
the counts' density lowers dramatically. As for $y'_{\text{{min}}}$,
the value of $250000$ in units of $\text{viral RNA copies/ml}$ indicates
the threshold for the isolation of infectious virus in cell cultures
at more than $5\%$ probability as described by Wörfel et al. \cite{munich}.
We then analyze the cobas and LC $480$ datasets first neglecting
the data below $y_{\text{min}}$ threshold and then below $y'_{\text{min}}$.
For the sake of simplicity, we will denote the cobas and LC $480$
datasets with $y_{\text{min}}$ and $y'_{\text{min}}$ as a lower
threshold as $d_{\text{C}}$, $d'_{\text{C}}$, $d_{\text{L}}$, and
$d'_{\text{L}}$, respectively. This way we can highlight the differences
between the two datasets and investigate possible sources of systematic
errors in the viral load measurement instruments or different selection
effects in the data collection process. 

The datasets are not explicitly provided by the authors. Therefore, we acquired 
the data by means of a plot digitizer algorithm from Fig.\ 6 of \cite{jones2020analysis}.
Since the age coordinates are not labeled precisely, but only a rough interval 
$\Delta_{\text{age}}\sim0-100\,\text{yr}$ is provided, we do not expect the acquired 
data points to be accurate -- especially in the age domain. For this reason the obtained age
axis could be affected by a global shift up to $5-10$ years in any
direction. But, since our model is translation invariant, this kind of systematic
bias in the data extraction process (an overall shift in the age values for all 
data points) does not affect the causal inference machinery. 
Nonetheless, the reader should keep this in mind when interpreting the results
concerning the conditional PDF of the viral load given the age. Concerning
the viral load coordinates, for which more precise units were given,
the data should be regarded as more reliable. For our aim of building
a method to analyze age and viral load data in a non-parametric and
causal fashion providing uncertainty estimates, this level of accuracy
is sufficient. The results we present from here on are given for the
measured coordinate values without considering any uncertainty with
respect to the real quantities (age, viral load). Nevertheless, we
do not believe systematic or random error contributions in the data
extraction procedure to significantly affect our results since the
shapes of the learned distributions are translation invariant.

\section*{Results\label{sec:Results}}

In the following, we discuss the main results and their relevance
for the infectivity of different age groups.

\subsection*{Age dependence of the viral load}

For both lower thresholds and datasets $d_{\text{C}}$, $d_{\text{L}}$,
we estimate the model parameters $s$ by means of the MGVI algorithm
implemented in \texttt{NIFTy}. In Figs.~\ref{fig:data-reconstruction} and 
\ref{fig:data-reconstruction-lc} we show the data (left panel) and the 
correspondent reconstructed underlying densities $\varrho(x,y)$ (central panel). 

\begin{figure*}
\centering
\includegraphics[width=1\textwidth]{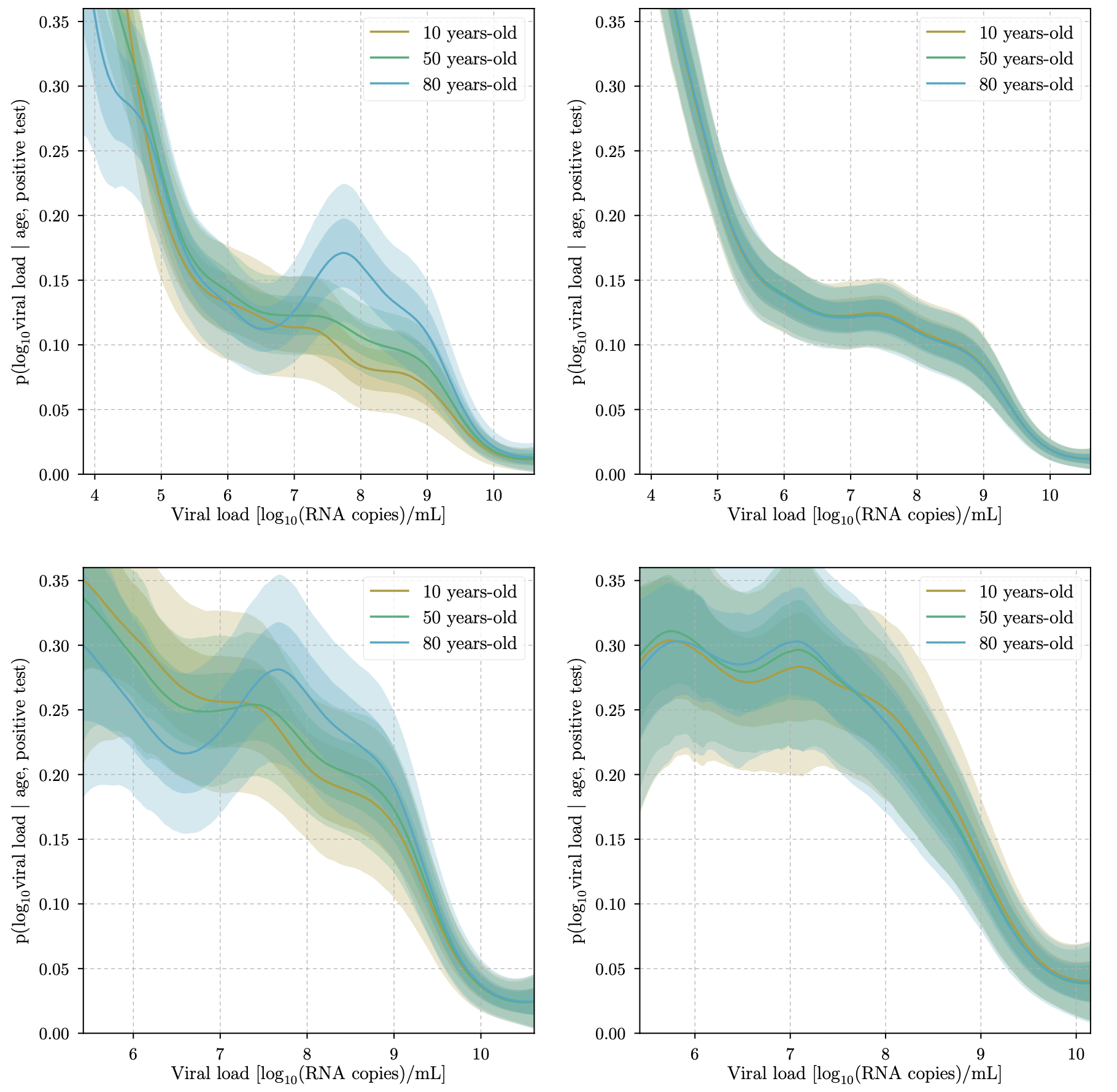}
\caption{\textbf{Viral-load probabilities for specific ages.} Viral load conditional 
PDF $p(y|x)$ for specific ages $x\in\{10,50,80\}$.
Panel one and two (top) display the results for the cobas dataset $d_{\text{C}}$ 
and for the randomized cobas dataset $d_{\text{C}}$ respectively.
The latter serves as a null test, since the randomization erases any (causal) 
relation between the age ($x$) and the viral load ($y$) other than shot noise. 
Panel three and four (bottom) show $p(y|x)$ for the cobas dataset $d'_{\text{C}}$ 
and for the LC $480$ dataset $d'_{\text{L}}$ respectively, both with the higher 
viral load threshold $y'_{\text{{min}}}$.
The shaded regions represent $1\sigma$ and $2\sigma$ uncertainty
contours of the approximate posterior.}
\label{fig:conditional-1}
\end{figure*}

A multi-modal age distribution is clearly visible as well as an overall
decrease for growing viral loads in all of the reconstructed densities.
The conditional PDF $p(y|x)$ of the viral load $y$ given the age
$x$ is shown in Fig.~\ref{fig:conditional-1}. In the conditional
probability, the multi-modal age structure displayed by the density
is not visible since it has been absorbed by $\varrho(x)$. What this
effectively means is that age-only selection effects -- e.g., testing
one or many specific age groups more than others or demographics in
general -- have been modeled, and the resulting conditional probability
distribution does not depend on such effects.

For all datasets and ages, a general descending trend in the viral
load probability distribution is clearly visible. The reconstruction
based on the $d_{\text{C}}$ dataset exhibits significant differences
in the viral load for different ages (Fig.~\ref{fig:conditional-1},
first panel). For infected patients approximately above the age of
$60$, the distribution exhibits a distinct maximum for viral loads
of $y\simeq8$ (in units of $\log_{10}(\text{RNA copies/ml})$). Furthermore,
we show that this feature is indeed triggered by the data, and is
not just the result of an over-fit of sample noise. To do so, we apply
a random permutation $r$ to the viral load values in the $d_{\text{C}}$
dataset, the only one that exhibits a possible ${x\rightarrow y}$
causal structure. We then analyze the randomized dataset ${d_{\text{C}}^{r}=\{(i,x_{i},y_{r(i)})\}_{i=1}^{N}}$
in the same way as seen for ${d_{\text{C}}=\{(i,x_{i},y_{i})\}_{i=1}^{N}}$.%
The resulting conditional PDF $p^{r}(y|x)$ reconstructed from the
randomized dataset (Fig.~\ref{fig:conditional-1}) does not exhibit
any clear age-dependent structure in the viral load, indicating that
the differences seen in the real data are not just a shot noise effect.

\subsection*{Causal directions and bias \label{subsec:causal_results}}

As discussed in Subsec.~\nameref{subsec:Causal-structure}, we can use 
ELBO ratios to identify causal relations. In our analysis, this corresponds to
choosing between the $x\rightarrow y$, $y\rightarrow x,$ and $x\perp y$ models. 
Identifying the causal direction underlying the data will allow to show that selection effects 
distorting the expected causal relation $x\rightarrow y$ are subdominant and 
that for the cobas dataset $d_{\text{C}}$ we indeed see evidence for an age dependence 
of the viral load distribution, $p(y|x)\neq p(y)$.

As a matter of fact, the log-evidence ratio for the $d_{\text{C}}$ dataset is $\Delta E_{\text{C},x\rightarrow y}=4.6\pm1.0$
for $d_{\text{C}}$, which clearly favors the dependent model, but
this value decreases to $\Delta E'_{\text{C},x\rightarrow y}=-1.5\pm1.0$
when considering $d'_{\text{C}}$, hence $y'_{\text{min}}$ as a lower
threshold. We highlight that a log-evidence difference between
the two compared models of $1$ unit corresponds to a factor of $e\approx2.7$
for the Bayesian odds ratio between the two. Thus, $\Delta E_{\text{C},x\rightarrow y}=4.6\pm1.0$ 
implies given equal model priors, $p(x\rightarrow y)=p(x\perp y)$,
a posterior model odds ratio of $p(x\rightarrow y|d):p(x\perp y|d)=e^{3.5\pm0.7}\approx99.5_{[36.6]}^{[270.4]}$
in favor of a causal dependence between viral load and age for the
cobas dataset with the lower threshold $y_{\text{min}}$.

For the opposite causal direction we get $\Delta E_{\text{C},y\rightarrow x}=-0.2\pm1.1$,
which shows that there is no strong $y\rightarrow x$ structure in
the data. For the LC dataset $d_{\text{L}}$ these evidence differences
with respect to the independent model become $\Delta E_{\text{L},x\rightarrow y}=-4.6\pm1.0$
and $\Delta E'_{\text{L},x\rightarrow y}=-3.4\pm1.0$ respectively
for the two thresholds. Therefore the independent model is favored
in both cases. This shows that the cobas ($d'_{\text{C}}$) and LC
$480$ ($d'_{\text{L}}$) datasets are in agreement for viral loads
which are higher than $y'_{\text{min}}$, but in case we include the
data lying in the viral load region $y\in[y_{\text{{min}}},y'_{\text{{min}}}]$,
the causal age-dependent structure becomes visible in $d_{\text{C}}$
and is not negligible anymore.

It is known that model evidences can vary strongly with different
data realizations. Moreover, in order to calculate the evidences,
we invoked approximations and stochastic calculation steps. Thus,
proper null-tests are required in order to validate and calibrate
the evidence ratio calculation. We provide such null-tests by repeating
the data randomization step described above several times, thereby
producing many randomized datasets. By construction, these dataset
should not exhibit any causal structure. Indeed, for the $10$ randomized-dataset
realizations performed, we find much lower log-evidence ratios between
dependent (causal) and independent models with respect to the ones
found for the original cobas dataset $d_{\text{C}}$. The average
difference between these tests is $\expval{\Delta E_{\text{random},
x\rightarrow y}}_{\text{randomizations}}=-6.0\pm1.0$.
Since none of the randomized dataset realizations reaches comparably
high log-evidence ratios with respect to the independent model, all
these findings support robustly the argument that the dependent model
is a more suited description of the $d_{\text{C}}$ dataset.
As previously discussed, this result also supports the 
assumption that no additional variable $z$ is confounding $x \leftarrow z \rightarrow y$
the bivariate causal direction $x \leftrightarrows y$.
Had there been a strong confounder $z$, we would have expected a more complicated 
data distribution that could have possibly been detected when comparing
 the actual data with the randomized datasets. 
 Even if this did not appear to be the case, we cannot completely rule 
 out the presence of a weak confounder.

\begin{figure*}[t]
\includegraphics[width=1.0\textwidth]{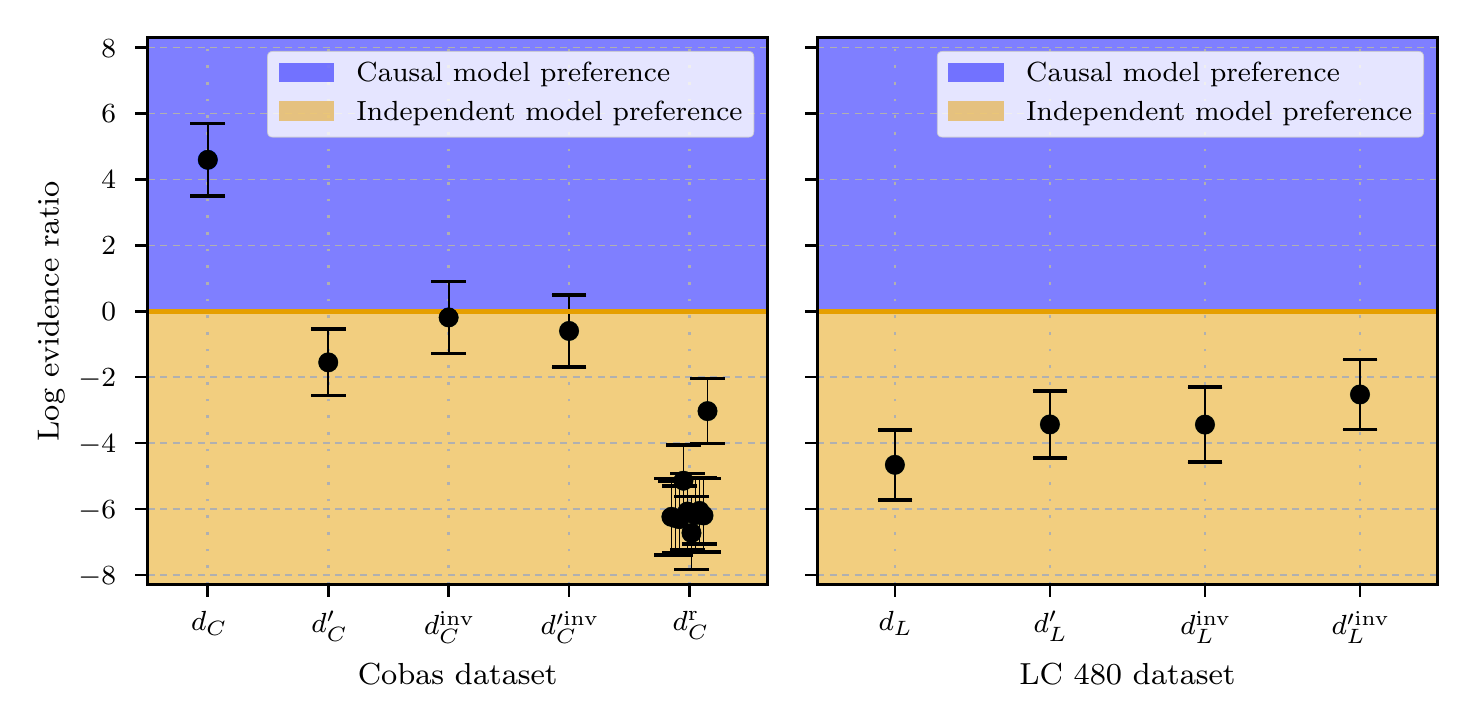}
\caption{\textbf{Evidence results.} Natural logarithm of the evidence ratio with respect to the correspondent independent
model for the $x\rightarrow y$ cobas datasets $d_{\text{C}}$ and
$d'_{\text{C}}$, for the $y\rightarrow x$ causal model $d_{\text{C}}^{\text{inv}}$
and $\left.d'\right._{\text{C}}^{\text{inv}}$(left panel).
The log-evidence ratios labeled with $d^r_C$ display the ($y$-)randomized datasets (for 10 different realizations).
The same is also shown for the LC $480$ dataset (right). The error bars represent 
the numerical uncertainty associated to the stochastic estimate of the ELBO. 
A positive logarithm of the evidence ratio denotes a preference for the given 
causal model with respect to the correspondent independent model and vice versa. 
The strength of this preference can be determined by taking the exponential of the log-evidence ratio 
between the causal model and the correspondent independent one. 
For example, a log-evidence ratio of $5$ corresponds to a $e^{5} \approx 141$-fold 
preference for the examined causal model. \label{fig:evidence_plots}}
\end{figure*}

The results of the evidence calculations are displayed in Fig.~\ref{fig:evidence_plots}.
This figure also indicates that the independent model interpretation
of the data is favored for all other datasets and threshold combinations
(except for $d_{\text{C}}$), since the evidence for the independent
model is always higher.

These contradicting results for the two datasets (or thresholds) $d_{\text{C}}$
and $d'_{\text{C}}$ might have several possible explanations. First,
they could be caused by a potential accuracy loss of the PCR devices
below certain viral load values, as suggested for the cobas dataset
$d_{\text{C}}$ below $y'_{\text{{min}}}$ in \cite{jones2020analysis},
hence for the Roche cobas 6600/8800 PCR system. It could also mean
that the opposite is happening and the Roche LC $480$ PCR device
is less sensitive than the Roche cobas 6600/8800 in the $y\in[3.8,5.0]$
region. This possibility would be supported by the fact that the cobas
dataset exhibits a clear age dependent structure in such viral load
region, but the (swab) data processed by the cobas PCR device contains
no information on the patients' ages. Hence it would be surprising
that a systematic effect in the measurements could introduce an age
dependence on the viral load distribution. Furthermore, this pattern
-- that older patients exhibit higher viral loads -- is plausible
from a medical perspective.

Nevertheless, we cannot exclude the possibility that selection effects
have been introduced in the data. We have already shown that ``viral
load causing age'' effects ($y\rightarrow x$) are subdominant. Nevertheless,
selection effects could still have been introduced for instance by
collecting age and viral load subsamples from a ``viral-load biased''
population sample. This could happen, e.g., if symptomatic patients
had been predominantly tested. Since children are less likely
to show symptoms than adults, the sample would then include mainly
those children with higher viral loads.

And finally, a combination of such competing effects could have affected
the results and thereby imprinted a spurious age dependence to the
viral-load distribution. Given only the statistical data in our possession,
this possibility cannot be ruled out completely.

\begin{figure*}[!t]
\centering
\includegraphics[width=1\textwidth]{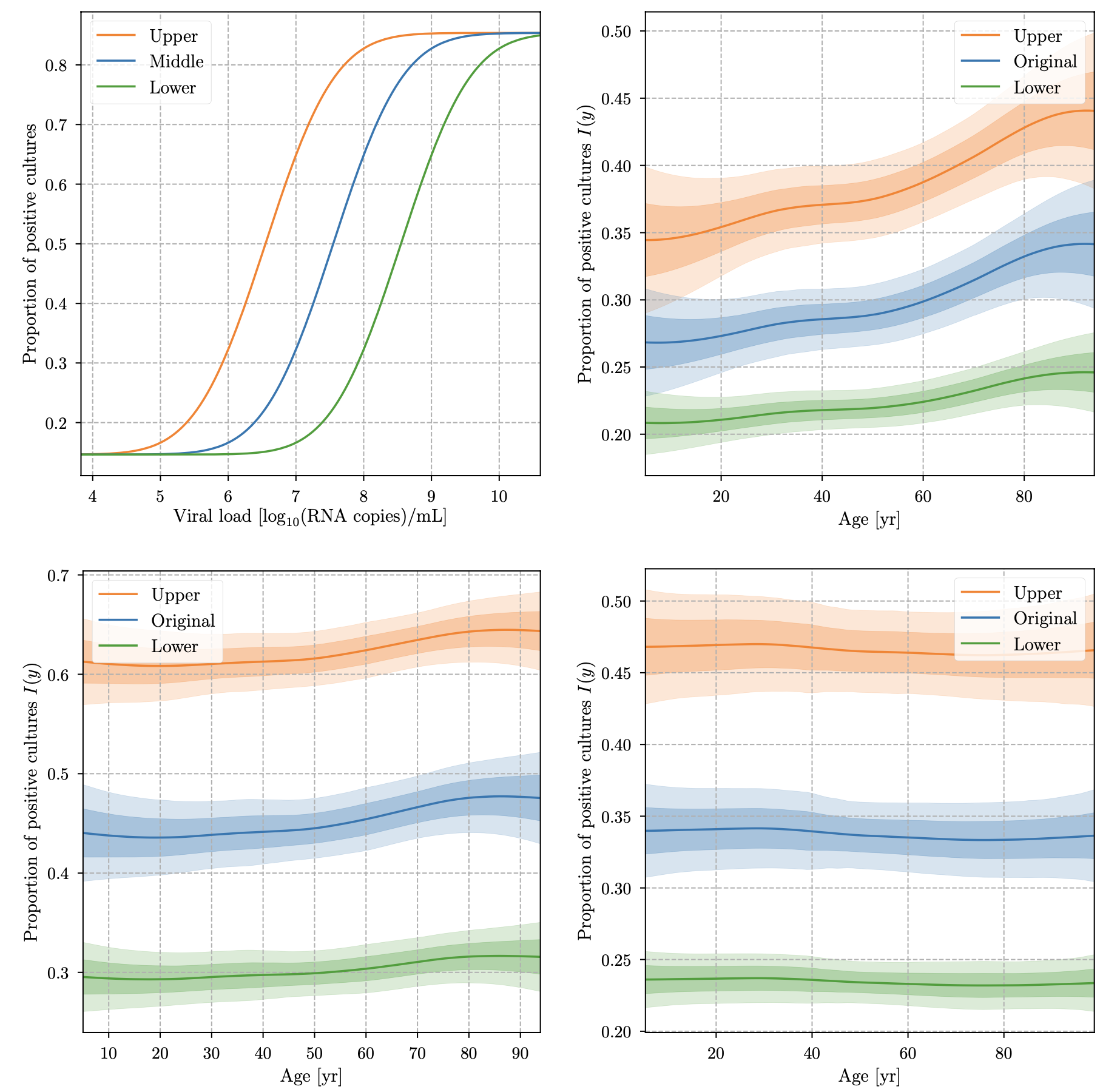}
\caption{\textbf{Infectivity proxy vs. age.} $I(y)$ obtained from Fig.~1g in \cite{munich} and fit with a probit
function (top left, Original), then projected for different patients'
ages for the cobas dataset $d_{\text{C}}$ (top right), $d'_{\text{C}}$
(bottom left) and for the LC $480$ dataset $d_{\text{L}}$ (bottom
right). The upper and lower curves are obtained by translating $I(y)$
of one unit in viral load as in $I(y-1)$ and $I(y+1)$ in order to
give upper and lower bounds similar as those shown in Fig.~1g in
\cite{munich}. The shaded regions show $1\sigma$ and $2\sigma$
uncertainty contours of the approximate posterior.\label{fig:infectivity}}
\end{figure*}

\subsection*{Impact on infectivity}

After having established a potential age difference in the viral load
distribution for $d_{\text{C}}$, we investigate whether this difference
- if real - would be relevant for the infection dynamics. For this
purpose, we need to link the viral load to the infectivity $I(y)$
of the virus, i.e. the probability of transmitting the infection.
Infectivity can be measured in different ways. In our analysis, we
choose the projected virus isolation success based on probit distribution
described in \cite{munich} as a proxy for infectivity. This represents
the infection success rate for cell cultures exposed to saliva with
different viral load $y$ and can be seen as the blue curve in Fig.~\ref{fig:infectivity},
labeled with ``Original''. We will from now on refer to this parameter
as ``infectivity proxy'' and indicate it with $I(y)$.

The projected infectivity as a function of the age is then given by
the expectation value of the infectivity parameter over the conditional
PDF $p(y|x)$ 
\begin{equation}
    I(x)\coloneqq \expval{I(y)}_{p(y|x)}=\int I(\tilde{y}) p(\tilde{y}|x) \dd{\tilde{y}}.
\end{equation}
The result, together with the uncertainties resulting from our PDF
modeling are shown in Fig.~\ref{fig:infectivity}. No relative differences
larger than \textbf{$0.18$} for the (projected) infectivity of the
different age groups is found, with typical values of $I(x)\approx0.3$
for all datasets.\textbf{ }This means that at most a $50\%$ difference
in infectivity due to different viral load between different age groups
- but more likely a smaller one - should be expected.

In order to characterize the uncertainty resulting from our $I(y)$
modeling, due to the uncertainty of the original determination of
this function and due to the uncertainty in the identification of
viral loads with different instruments, we repeat the analysis while
shifting the original $I(y)$ curve by one order of magnitude upwards
and downwards in $y$. The resulting maximal relative difference in
the infectivity of the different age groups is $\simeq0.3$. Thus,
even though our model allows us to show that infectivity exhibits an
age dependence if the \textbf{$d_{\text{C}}$ }dataset with $y_{\text{{min}}}$
provides a valid picture, the viral load differences between different
age groups though are not strong enough to impact on the infection
dynamics at a level that justifies regarding any age group as noninfectious
or even significantly less infectious.

\section*{Conclusions}
\label{sec:Conclusions}

In order to investigate the controversial results reported
in the literature, we developed a causal model to assess the dependence
of viral loads of patients infected with COVID-19 on age. The developed 
model is capable of reconstructing two-dimensional density distributions 
from data counts and to learn causal directions. The model complexity is set
in a user-independent fashion making the results more robust and consistent.
Furthermore, its causal nature allows to make predictions about non-directly measured 
quantities (e.g. the infectivity assessment in Sec.~\nameref{sec:Results} and 
Fig.~\ref{fig:infectivity}) and to additionally test for bias in the data-collection process.

Although the benefits of a causal analysis were already discussed in Pearl's
original work \cite{reason:Pearl09a} and are usually well recognized, 
causal inference is not so commonly applied to real-world data science,
often because it requires the implementation of complicated ad-hoc models. 
With the hope of making it useful for a wide range of data-science applications we
make our causal model freely available.
This model is very flexible and generic (as it only requires a set of $x$ and $y$ data pairs
and mild priors on their correlation structures) and it can therefore be used in future epidemiological
studies as well as in completely different fields. 
We provide the source code under an open source license for usage in further studies and applications 
at \href{https://gitlab.mpcdf.mpg.de/ift/public/causal_age_viral_load_model}
{https://gitlab.mpcdf.mpg.de/ift/public/causal\_age\_viral\_load\_model}. 
As a side product, we also developed a causal-direction-agnostic density 
estimator, which is described in more detail in the \nameref{sec:Mat=0000E9rn-kernel-density}.

Using our novel method to model causal relations non-parametrically,
we have reanalyzed the SARS-CoV-2 age and viral load data presented
in \cite{jones2020analysis}.In doing so, we have found
statistically significant differences in the viral load distribution
of different age groups when regarding the cobas dataset $d_{\text{C}}$
for viral loads within the interval of $10^{3.8}$ to $10^{5.0}$
in units of $\text{viral RNA copies/ml}$ of sample or entire swab
specimen as reliable. These differences become irrelevant
if this region is ignored in the analysis.

We cannot completely exclude that selection effects in the data-collection
process may have introduced an apparent causal relation between viral
load and age, but the observed trend -- a statistically-significant
increase in the viral load with age -- fits with the generally accepted
notion that the immune system response gets weaker with age. Assuming
this trend to be real, we showed, however, that its expected impact
on the infectivity of different age groups is at most moderate. For
this reason we cannot exclude any age group from being considered
as a potentially significant source of infection.

The region of the cobas dataset relevant for this trend is described
in \cite{jones2020analysis} as containing an artifact, suggesting
that the correct interpretation of the data is that viral load, hence
infectivity, is predominantly age independent. Here, we want to point
out that other studies on the age dependence of the viral load present
in the literature \cite{Euser2021.01.15.21249691,10.3389.fmed.2021.608215}
make the opposite claim. Moreover, in their most recent publication, 
Jones et al. \cite{Joneseabi5273} acknowledge an age dependence of the viral load. 
This dependence is quantitatively similar to the one we have detected 
with our method. Furthermore, the causal evidence tests presented
in Sec.~\nameref{sec:Results} favor considering the age dependence
of the viral load as a real effect and not just as an artifact. 
These tests also disfavor the reverse-causal-direction model (here: that
the viral load of a patient \textquotedblleft causes\textquotedblright{}
its age), which would indicate that strong selection effects have
affected the data-collection process. We introduced these tests as
a new tool to detect potential systematic effects in similar datasets. 

In conclusion, the results of our analysis ultimately confirm 
some of the findings in the literature - i.e. that the viral load is only
modestly dependent on the age - but with a much higher sensitivity and 
robustness. Of central importance are the methods here developed.
While being tailored to describe the Covid-19 pandemic data, they
can be easily adapted for more general purposes and can prove 
very useful also for future pandemics or for new -- and possibly more 
infective -- mutations of SARS-CoV-2.

\section*{Supporting information}

% Include only the SI item label in the paragraph heading. Use the \nameref{label} command to cite SI items in the text.

\paragraph*{S1 Appendix. Matérn-kernel density reconstruction.} 
\label{sec:Mat=0000E9rn-kernel-density}

{\bf Matérn-kernel density reconstruction.} We aim to infer an unknown distribution from a random realization
of discrete data. In the following, we present a general-purpose density
estimator that serves this scope. Using a fully Bayesian framework,
we can propagate uncertainties through each inference step and extract
information from the correlation structure inherent to the data. We
provide our method as open-source software.

According to Chen, Tareen, and Kinney \cite{density}, the problem of extracting a smooth
density function from a limited set of data samples is a challenging
and well-known problem in statistical learning and data analysis.
The most common ad-hoc methods to empirically derive (probability)
densities from data usually involve histograms or Kernel Density Estimation
(KDE) \cite{silverman1986density,sheatherdensity}. These methods
do not infer the smoothness of the learned (probability) density's
correlation structure and are thus prone to reconstructing unphysical
densities. Other methods make use of neural networks (see for example
Liu et al. \cite{Liue2101344118}) or restrict the density to specific functional
forms (see e.g. Dirichlet Process Mixture Model, or DPMM 
\cite{ferguson,mueller2015bayesian,gelman2014bayesian}).
Another approach is to use smooth priors and infer the level 
of smoothness of the reconstructed density from data via maximum entropy 
(Density Estimation using Field Theory, or DEFT \cite{PhysRevE.90.011301,PhysRevE.92.032107}).
Chen, Tareen, and Kinney \cite{density} propose an interesting information-theoretical-based 
modification of DEFT, although it only effectively works in one dimension.
Finally, another very commonly used and effective solution to the density 
estimation problem is the one given by deep neural networks. In a recent work, 
Liu et al. \cite{Liue2101344118} propose a generative adversarial networks (GAN) 
which is particularly effective in high dimensions. We refer to their work 
for comparison with similar neural-network-based approaches.

Most of these approaches lack a robust estimate for uncertainties or 
specify none at all. In the hope of addressing these shortcomings, 
we propose our novel general-purpose density reconstruction method,
which we will refer to as Matérn Kernel Density Estimator (MKDE).
This method is very general, works for a generic $n$-dimensional
space, and therefore applies to many different contexts and fields.
We presented one paradigmatic example of its many possible applications
in Sec.~\nameref{subsec:Causal-structure}. In this example, we used
MKDE to reconstruct the continuous distributions of the ages and viral
loads of patients infected with Covid-19 from age-and-viral-load data
samples collected within the general population.

MKDE is capable of reconstructing a smooth density distribution underlying
an -- even limited -- discrete dataset. We achieve this result under
the hypotheses that the data points are drawn from the underlying
density through a Poisson process and that the reconstructed density
is a sufficiently smooth function. Since we expect the inferred (probability)
density to be strictly positive and to vary on logarithmic scales,
we choose the log-normal model
\begin{equation}
\varrho(x)=e^{s(x)},\label{eq:LogNormalDensity}
\end{equation}
where $s(x)$ is the natural logarithm of the density. In the case
of multi-dimensional data, both the density $\varrho(x)$ and its
logarithm $s(x)$ in Eq.~\ref{eq:LogNormalDensity} are functions
defined over $n$-dimensional vector spaces, with $n$ being the dimension
of the data space (e.g. space, time, age, viral load, ...). For the
sake of simplicity, we will initially show how a one-dimensional density
is reconstructed, emphasizing how to generalize to the multi-dimensional
case only when this generalization is non-trivial.

\begin{figure*}[t]
\includegraphics[width=1\textwidth]{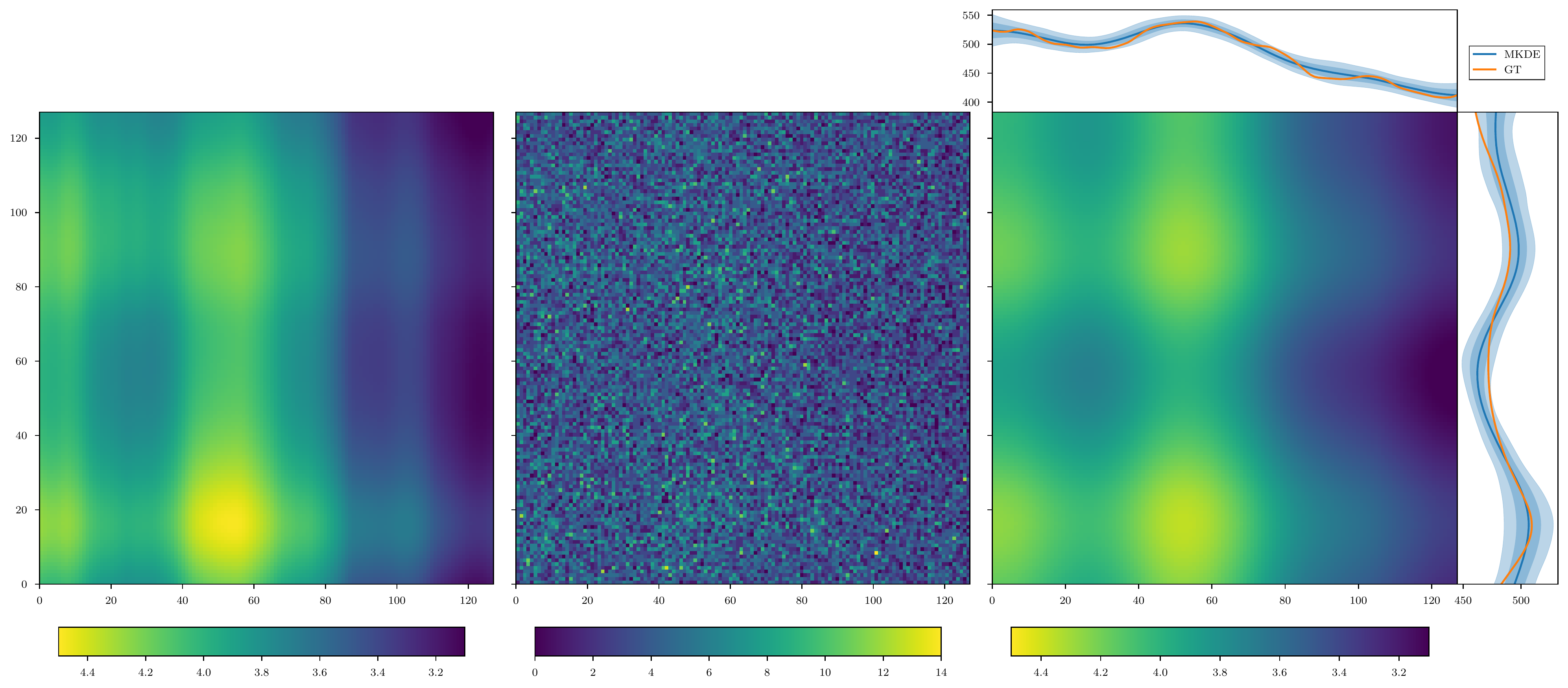}

\caption{\textbf{ Example of MKDE performance.} Left: random realization of a smooth
probability density with independent covariance structure in the $x$
and $y$ direction (ground truth, GT). Middle: Poissonian counts drawn
from the GT. Right: MKDE reconstruction from the counts. In the same
panel, we also show the marginals of the reconstructed density (blue
lines), together with their one and two-sigma uncertainty estimates
(shaded areas in blue) against the marginals of the GT (orange lines).\label{fig:Example-of-MKDE}}
\end{figure*}

Smoothness is a common and ubiquitous assumption when dealing with
physical data. To fulfill this assumption in the MKDE, we parametrize
the two-point correlation structure of the Gaussian process that determines
the value of the log-density $s(x)$ at each point and for each dimension
with a Matérn kernel. This kernel is a very flexible choice \cite{matern}.
Moreover, it is very well suited to represent a priori homogeneous
covariance structures like the ones we want to model. We define the
Gaussian process $s:=s(x)=(s_{x})_{x\in\mathbb{R}}\hookleftarrow\mathcal{P}(s)=\mathcal{G}(s,S)$
on the one-dimensional position space $\mathbb{R}$. This process
has a homogeneous covariance structure $S_{xx'}=C_{s}(x-x')$ that
can be efficiently represented in Fourier space thanks to the Wiener-Khinchin
theorem. Invoking this theorem, we can use the power spectrum $P_{s}(k)$
to fully determine the Fourier transform of the two-point correlation
function $C_{s}(x-x')$ for any stationary and statistically homogeneous
process, and in particular for the Gaussian process $s$.

In order to draw prior samples from the Gaussian field $s$, we choose
a standardized coordinate system $\hat{\xi}$. We then transform the standard 
normally distributed parameters $\xi=(\xi_{k})_{k}$, where $k\in\mathbb{N}$ is the Fourier-space
index according to the mapping
\begin{equation}
    s=\mathcal{F}^{-1}\,A\,\xi\quad\text{with }A\coloneqq\text{diag}(\sqrt{P_{s}}).
\end{equation}
Here, $\mathcal{F}$ represents the Fourier transform operator and
$A$ the amplitude operator in Fourier space. The amplitude operator
encodes the Matérn-kernel correlation structure, parametrized with
\begin{equation}
    A_{kk'}=\,2\pi\,\delta(k-k')\frac{a_{s}}{\left[1+k^{2}/k_{0}^{2}\right]^{-\gamma_{s}/4}},\label{eq:MaternAmplitude}
\end{equation}
where $a_{s}$ is a scale factor which accounts for the standard deviation
in position space, $k_{0}$ is the magnitude of the characteristic
correlation-length wavevector, and $\gamma_{s}$ is the spectral index
of the power spectrum. We assume $a_{s}$ and $k_{0}$ to be a priori
log-normally distributed since we expect strictly positive variations
of the possible power spectra on a logarithmic scale. Similarly, we
choose $\gamma_{s}$ to be normally distributed, since the spectral
index could in principle also be negative. We additionally introduce
volume factors to ensure that the model parameters are intensive with
respect to volume, i.e. they do not depend on the volume in position
space.

For higher-dimensional data, we expect the correlation structure along
each axis (or dimension) to be a priori independent from the others.
These different axes could in fact have very different meanings (and
units), as they could represent -- for instance -- space and time,
temperature, pressure, and volume, age and viral load (as seen in
Sec.~\nameref{subsec:Causal-structure}), or a different combination
of these and other continuous quantities. Therefore, in the $n$-dimensional
data space we can decompose the amplitudes of the correlation structure
\begin{equation}
A_{k,q,\ldots,n}=\bigotimes_{i\in\{k,q,\ldots,n\}}A_{i}\label{eq:outer-product}
\end{equation}
along each independent axis, each modeled by an individual amplitude
operator $A_{i}$, for $i\in\{k,q,\ldots,n\}$. The zero modes of
the individual axes must be treated separately in order to avoid degeneracy.
Thus, in the proposed model the zero mode is shared among all directions
and inferred independently through an a priori strictly-positive and
uniformly-distributed parameter $\alpha$. For more details on the zero mode 
degeneracy and factorizing power
spectra, we refer to Arras et al. \cite{2021m87}.

At this stage, we can summarize all the parameters that we have introduced
for each independent axis with the scalar-valued parameters $\alpha,a_{s},k_{0}$,
and $\gamma_{s}$ and the vector-valued $\xi_{k}$. We set broad priors
on these parameters and learn them using MGVI.
We set the following priors on the signal parameters: $\alpha=[10^{-15},5.0]$,
$a_{s}=(0.3\pm0.2)$, $k_{0}=(4.0\pm3.0)$, and $\gamma_{s}=(-6.0\pm3.0)$, 
where the mean and standard deviation specify a Gaussian prior distribution for 
$\gamma_s$ and log-normal distributions with the given mean and standard 
deviation for $a_s$ and $k_0$.
For details on the inference of posterior estimates for the MKDE
parameters through MGVI and their uncertainty quantification, we refer
to Sec.~\nameref{subsec:Inference}. Fig.\ \ref{fig:Example-of-MKDE}
illustrates the performance of MKDE in a two dimensional setting.

In conclusion, we described MKDE, a Matérn-kernel-based, Bayesian,
and non-parametric density estimator that can construct a smooth (probability)
density function from an - even limited - set of data samples.
The broad priors on the learned parameters, combined with the log-normal
model and the Matérn kernel covariance structure, make MKDE very flexible
and robust. Furthermore, the Bayesian inference framework allows for
posterior uncertainty quantification for the reconstructed density.
A software implementation of MKDE is available in NIFTy 7 and is also
released as an open-source Python package (DENSe), which can be found
at: \href{https://ift.pages.mpcdf.de/public/dense/}{https://ift.pages.mpcdf.de/public/dense/}.

\paragraph*{S2 Appendix.}
\label{S2_Appendix}
{\bf Priors.} 
Throughout the analysis, we have made the use of the following priors on the signal parameters: 
${a_f = (0.3\pm 0.1)}$, ${k_f=(5.0\pm3.0)}{\ \text{yr}^{-1}}$, and ${\gamma_f=(-3.0\pm2.12)}$, where 
the mean and standard deviation specify a Gaussian prior distribution for $\gamma_f$ 
and log-normal distributions with the given mean and standard deviation for $a_f$ and $k_f$.

\section*{Acknowledgments}
We thank Dr. Matthia Sabatelli (Montefiore Institute, University of Liège) for feedback on 
preliminary versions of this work and the Information Field Theory group at the Max Planck 
Institute for Astrophysics for the fruitful discussions.
\nolinenumbers

% Either type in your references using
% \begin{thebibliography}{}
% \bibitem{}
% Text
% \end{thebibliography}
%
% or
%
% Compile your BiBTeX database using our plos2015.bst
% style file and paste the contents of your .bbl file
% here. See http://journals.plos.org/plosone/s/latex for 
% step-by-step instructions.
% 

\end{document}